\RequirePackage{fix-cm}
\documentclass[smallextended]{svjour3}       
\smartqed  
\usepackage{graphicx}
\usepackage{epsfig}\usepackage{latexsym}\usepackage{amssymb}\usepackage{bm}
\usepackage{float}
\usepackage{subfigure}
\usepackage{amsmath}
\usepackage{amsfonts}
\usepackage[usenames]{color} 
\newcommand{\be}{\begin{equation}} \newcommand{\ed}{\end{displaymath}}
\newcommand{\bd}{\begin{displaymath}} \newcommand{\ee}{\end{equation}}
\newcommand{\bea}{\begin{eqnarray}} \newcommand{\eea}{\end{eqnarray}}
\newcommand{\ba}{\begin{array}} \newcommand{\ea}{\end{array}}

 \journalname{International Journal of Theoretical Physics}
\begin{document}

\title{Hofstadter's Cocoon}


\author{Katherine Jones-Smith  \and
        Connor Wallace 
}


\institute{K.Jones-Smith \at
Physics \& Astronomy Department \\
Oberlin College\\
110 N. Professor Street\\
Oberlin, OH 44074\\
U.S.A.
              Tel.: +440-775-6730\\
              \email{kjonessm@oberlin.edu}           
           \and
           C.Wallace \at
              Reed College 
              Portland, OR 97202}

\date{Received: 23 March 2014 / Accepted: 29 May 2014}

\maketitle

\begin{abstract} Hofstadter showed that the energy levels of electrons on a lattice plotted
as a function of magnetic field	form an beautiful structure now referred to as
``Hofstadter's butterfly''. We study a non-Hermitian continuation of
Hofstadter's model;
as the non-Hermiticity parameter $g$ increases past a sequence of critical values
the eigenvalues successively go complex in a sequence of ``double-pitchfork bifurcations''
wherein pairs of real eigenvalues degenerate and then become complex conjugate pairs.
The associated wavefunctions undergo a spontaneous symmetry
breaking transition that we elucidate.
Beyond the transition a plot of the real parts of the eigenvalues against magnetic field
resembles the Hofstadter butterfly; a plot of the imaginary parts plotted against magnetic
fields forms an intricate structure that we call the Hofstadter cocoon. 
The symmetries of the cocoon are described.
Hatano and Nelson have studied a non-Hermitian continuation of the Anderson model
of localization that has close parallels to the model studied here. 
The relationship of our work to that of Hatano and Nelson
and to PT transitions studied in PT quantum mechanics is discussed. 

\end{abstract}
\maketitle

\begin{abstract} Hofstadter showed that the energy levels of electrons on a lattice plotted
as a function of magnetic field	form an beautiful structure now referred to as
``Hofstadter's butterfly''. We study a non-Hermitian continuation of
Hofstadter's model;
as the non-Hermiticity parameter $g$ increases past a sequence of critical values
the eigenvalues successively go complex in a sequence of ``double-pitchfork bifurcations''
wherein pairs of real eigenvalues degenerate and then become complex conjugate pairs.
The associated wavefunctions undergo a spontaneous symmetry
breaking transition that we elucidate.
Beyond the transition a plot of the real parts of the eigenvalues against magnetic field
resembles the Hofstadter butterfly; a plot of the imaginary parts plotted against magnetic
fields forms an intricate structure that we call the Hofstadter cocoon. 
The symmetries of the cocoon are described.
Hatano and Nelson have studied a non-Hermitian continuation of the Anderson model
of localization that has close parallels to the model studied here. 
The relationship of our work to that of Hatano and Nelson
and to PT transitions studied in PT quantum mechanics is discussed. 

\end{abstract}
\maketitle

In 1976 Hofstadter demonstrated that the classic Landau problem, describing
an electron on a 2-D surface with a normal magnetic field, has a much more
interesting eigenvalue spectrum when solved on a lattice than in the 
continuum \cite{hofstadter}. For the lattice, a plot of energy levels as a function of
magnetic field forms an intricate
and beautiful pattern known as Hofstadter's butterfly, whereas in the 
continuum the energy levels disperse linearly with magnetic field,
forming the much simpler ``Landau fan''. A similar butterfly emerges
in the continuum when the Landau levels are weakly perturbed by a periodic
potential. Until recently the butterfly had eluded experimental observation
due to the unattainably high magnetic fields needed to insert a flux quantum
through an atomic scale unit cell. However an electromagnetic analog of 
Hofstadter's butterfly was observed using guided microwaves \cite{microwave}. Very recent
experiments have created a Moire superlattice by placing bilayer graphene
on a suitable substrate resulting in a potential with a periodicity of 
hundreds of Angstrom  \cite{exp1},\cite{exp2}, \cite{exp3}. These experiments have finally observed the butterfly
in the original context of electrons in a perpendicular magnetic field.
 
In this work we show that the Hofstadter model is the Hermitian limit of a more
general non-Hermitian Hamiltonian. In the Hermitian limit the spectrum of the
Hofstadter model is real.
But as the non-Hermiticity parameter parameter $g$
increases past a sequence of critical values the eigenvalues successively
go complex in pairs. As $g$ passes through a critical value a pair of real
eigenvalues degenerates and thereafter becomes a conjugate pair of 
complex eigenvalues. At the critical value the corresponding eigenfunctions
undergo a spontaneous symmetry breaking transition. Our study of the
evolution of the spectrum with $g$ reveals a rich and intricate pattern of
symmetry breaking transitions.  
These transitions can be visualized by plotting the real and imaginary
parts of the eigenvalues against the magnetic field. The plots of the
real parts resemble the Hofstadter butterfly; we dub the plots of the imaginary
parts the Hofstadter cocoon. Apart from its intrinsic interest, the behavior
that we calculate is experimentally accessible to microwave experiments \cite{microwave}
similar to those first used to observe the Hofstadter butterfly.  Recently Yuce has considered the 
same system in the context of coupled optical waveguides \cite{yuce}; his
findings are highly relevant and complementary to the work presented here.

Non-Hermitian continuations of Hermitian models have been fruitfully 
studied in condensed matter physics since the seminal work of Dyson 
on spin-waves in a ferromagnet \cite{dyson}. Our work has a close relationship to
that of  Hatano and Nelson,  who studied a non-Hermitian Hamiltonian that 
described the classical statistical mechanics of vortex line depinning \cite{nelson1} \cite{nelson2}, 
\cite{nelson3}.
The model of Hatano and Nelson may be regarded as  the non-Hermitian
continuation of the Anderson model of localization \cite{anderson}.
Hatano and Nelson made the interesting discovery that in their model,
the transition
to complex eigenvalues was accompanied by a delocalization transition
in the associated wavefunctions. Shortly thereafter, Bender and co-workers
initiated the study of non-Hermitian Hamiltonians with PT symmetry \cite{bender1} \cite{bender2}. 
Within PT quantum mechanics the transition to complex eigenvalues may be
regarded as the spontaneous breaking of PT symmetry. Interest in this
remarkable new kind of symmetry breaking has been heightened by the observation
of such a transition in optical systems with PT symmetry \cite{opt1}, \cite{opt2}. 
We show below that the transitions we study as well as the delocalization
transition of Hatano and Nelson are analogs of the PT transition inasmuch 
as they involve the spontaneous breaking of an anti-unitary symmetry. 

To provide context for our results we briefly recall the  
Hosfstadter and  Hatano/Nelson models. In the continuum, electrons in
the $x$-$y$ plane subject to a perpendicular magnetic field
${\mathbf B} = B \hat{{\mathbf z}}$ are governed by the
Schr\"{o}dinger equation
\begin{equation}
- \frac{1}{2} \left[ \frac{\partial^2}{\partial x^2} + \left(
\frac{\partial}{\partial y} - i x \right)^2 \right] \psi = E \psi.
\label{eq:landau}
\end{equation}
We have adopted the Landau gauge ${\mathbf A} = B x \hat{{\mathbf y}}$
and units wherein $\hbar = m = eB = 1$. 
We take the solution to be separable $\psi(x,y) = \xi(x) \exp( i k y)$;
then $\xi$ obeys
\begin{equation}
\left[ - \frac{1}{2} \frac{\partial^2}{\partial x^2} + \frac{1}{2} (x - k)^2 \right] \xi
= E \xi.
\label{eq:oscillator}
\end{equation}
This is a shifted harmonic oscillator; the energy levels of the system
are given by $E_n = (n + \frac{1}{2}) \hbar e B/m$. Thus we see that the energy
levels are evenly spaced with a spacing proportional to the magnetic field,
leading to the famous Landau fan. Note that if restrict the electron to
a square region of size $L$ and impose periodic boundary conditions
the allowed $k$ values become quantized but the eigenvalues are not
affected \cite{halperin}.

Hofstadter considered electrons on a two dimensional square lattice
of lattice constant $a$ immersed in a uniform perpendicular field \cite{hofstadter}. 
Thus $x = na$ and $y = ma$ and the 
Schr\"{o}dinger equation takes the form
\begin{equation}
- \tau e^{i 2 \pi \phi m} \psi_{n+1,m} 
- \tau e^{-i 2 \pi \phi m} \psi_{n-1,m} 
- \tau \psi_{n, m+1}
- \tau \psi_{n, m-1} 
 =  E \psi_{n,m}
\label{eq:hofstadter}
\end{equation}
where $\phi = e B a^2/h$ is the magnetic flux per plaquette (in units of $h/e$)
and we are working with a linear Landau gauge for the lattice model as well.
If we consider a finite system of size $a L \times a L$ 
and impose periodic boundary conditions then Dirac's quantization
argument requires that the total flux through the sample be an
integer or half integer. 
Periodic boundary conditions are the requirement that the wavefunction
is invariant under magnetic translations (rather than ordinary translations)
by $a L$ along the $n$ and $m$ directions \cite{duncan}. In the linear Landau gauge this
amounts to the conditions $\psi_{n+L, m} = \psi_{n,m} e^{ - i 2 \pi \phi L n }$
and $\psi_{n, m+L} = \psi_{n,m}$. Note that for the special case that the flux
obeys the condition that $\phi L$ is an integer the magnetic periodic boundary
conditions coincide with ordinary periodic boundary conditions $\psi_{n+L, m} = \psi_{n,m}$
and $\psi_{n, m+L} = \psi_{n,m}$. 
For simplicity we will restrict attention to these special values of flux for which the solutions to 
(\ref{eq:hofstadter}) have the separable form
\begin{equation}
\psi_{n,m} = e^{i k n} \xi_m,
\label{eq:latticeseparate}
\end{equation}
where $\xi_m$ obeys Harper's equation,
\begin{equation}
- \xi_{m+1} - \xi_{m-1} - 2 \cos (2 \pi \phi m + k) \xi_m = E \xi_m,
\label{eq:harper}
\end{equation}
with the boundary condition that $\xi_{m+L} = \xi_m$. 
Periodic boundary conditions also constrain $k$ to be of the form
$2 \pi p/L$ where $p$ is an integer. 
The eigenvalues of Harper's equation (\ref{eq:latticeseparate})
for all allowed values of $k$ plotted as a function of $\phi$ 
form the famous butterfly. In the infinite size limit the discrete
levels of the butterfly blend to form bands with a intricate pattern
of gaps. The self-similar structure of these bands was later
elucidated by Wannier \cite{wannier} and MacDonald \cite{macdonald}; Harper's equation 
also appears in the context of quasicrystals, see for example \cite{vincenzo}. 

Hatano and Nelson studied a one dimensional lattice model 
governed by
\begin{equation}
- e^{g} \xi_{m+1} - e^{-g} \xi_{m-1} - V_m \xi_m = E \xi_m
\label{eq:nelson}
\end{equation}
with the onsite potentials $V_m$ taken to be random \cite{nelson1}, \cite{nelson2}, \cite{nelson3}. 
Note that in place of phase factors that represent abelian gauge
fields on a lattice, Hatano and Nelson inserted real exponential
factors in the hopping terms corresponding to an imaginary vector potential. 
For $g=0$ the Hatano and Nelson model coincides with the
Anderson model of localization theory \cite{anderson}; 
for non-zero $g$ it corresponds to a non-Hermitian continuation
of the Anderson model. Hatano and Nelson found that for small
values of $g$ the eigenvalues remained real but as $g$ passed
a critical value the eigenvalues become complex in conjugate pairs.
Moreover the associated eigenfunctions underwent a delocalization
transition. 

Here we study the Hofstadter model perturbed by an imaginary
vector potential; a non-Hermitian
continuation of Harper's equation:
\begin{equation}
- e^{g} \xi_{m+1} - e^{-g} \xi_{m-1} - 2 \cos (2 \pi \phi m + k)  \xi_m = E \xi_m.
\label{eq:cocoon}
\end{equation}
The exponential factors corresponding to an imaginary vector potential differentiate
our model from the Hermitian Harper's equation. Our model differs from Hatano and Nelson in that the onsite potential is quasi-periodic rather than random.

\begin{figure}[h]
\begin{center}
\includegraphics[width=0.7\textwidth]{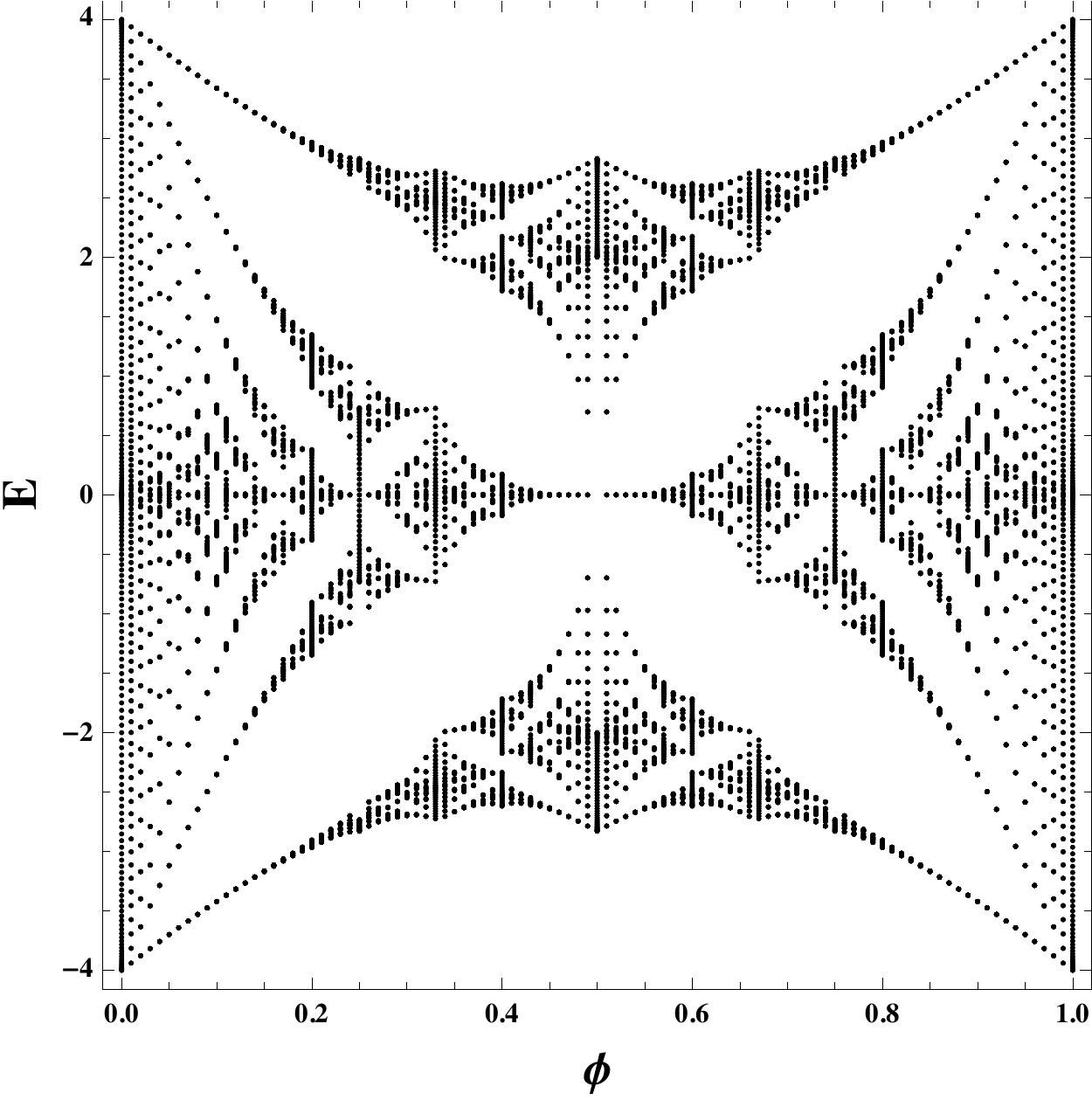}
\end{center}
\caption{{\em Hofstadter's butterfly.} Eigenvalue spectrum of Eq.(\ref{eq:cocoon}) plotted of as a
function of flux $\phi$ for $g = 0$ and $L = 200$.}
\label{fig:butterfly}
\end{figure}
\begin{figure}[h]
\begin{center}
\;\;\; \includegraphics[width=0.45\textwidth]{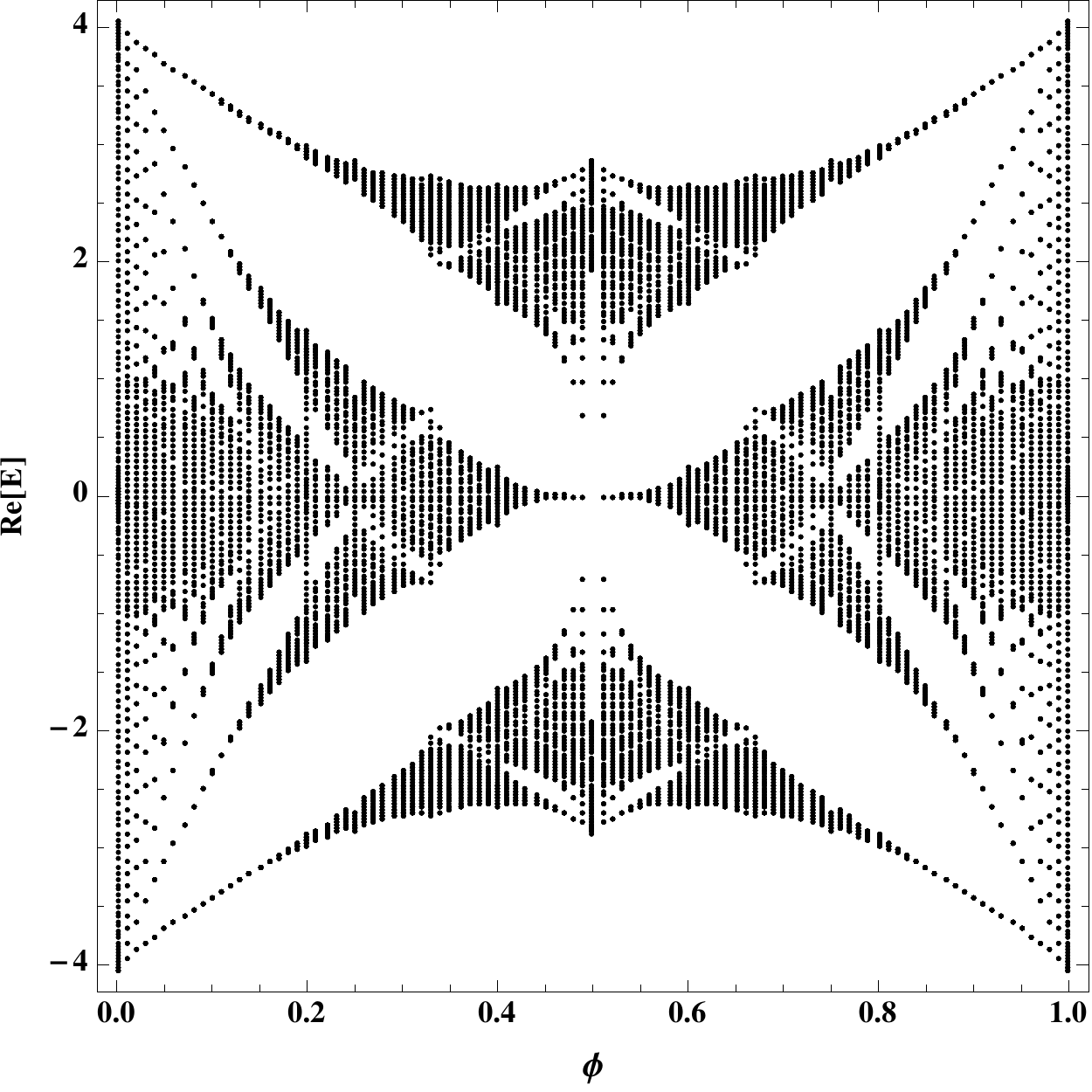}\;\;\;\;
\includegraphics[width=0.45\textwidth]{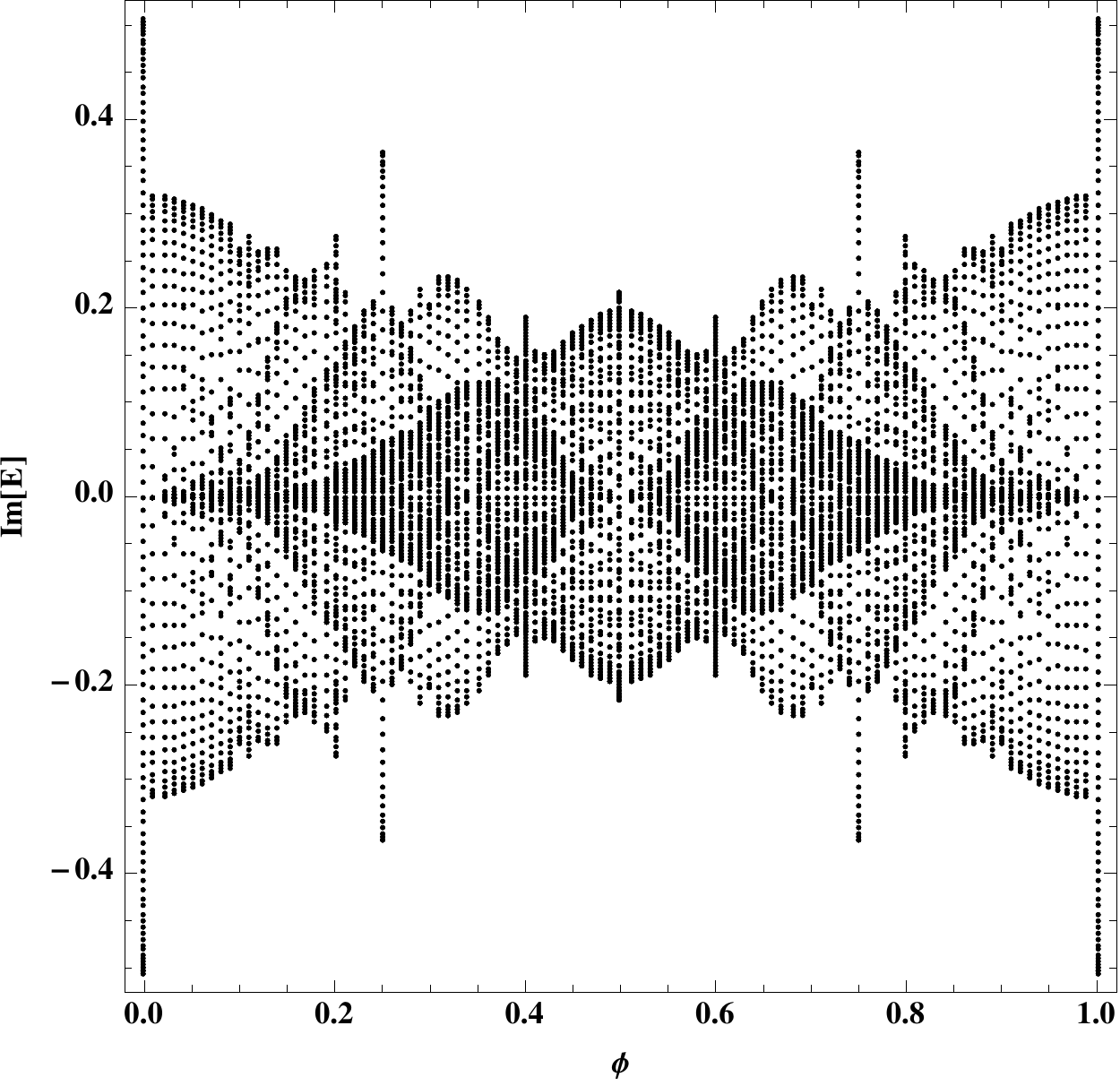}
\end{center}
\caption {The non-Hermitian butterfly and cocoon. Real (left) and imaginary (right) parts of the
eigenvalues of Eq.(\ref{eq:cocoon}), plotted as a function of flux $\phi$
for $g = -0.25$ and $L = 200$. }
\label{fig:nonhermbutterfly}
\end{figure}

Figure 1 shows the Hermitian Hofstadter butterfly: with $g=0$ the Hamiltonian in Eq. (\ref{eq:cocoon}) is Hermitian and the eigenvalues are guaranteed to be real. Figure 2 depicts
the spectrum for $g = -0.25$; . The left panel 
shows the real parts of the eigenvalues plotted as a function of flux. This plot,
which we call the non-Hermitian butterfly, resembles the Hofstadter butterfly qualitatively, although 
there are small quantitative differences that evolve with $g$. The right panel shows the imaginary
parts of the eigenvalues plotted as a function of flux, which form a complex structure we dub the
Hofstadter cocoon. In Figure 3 we illustrate the double pitchfork morphology of the transition to
complex eigenvalues that is characteristic of our model as well as of the model of Hatano and 
Nelson \cite{nelson3} and of the transitions studied in PT quantum mechanics \cite{bender2}. 
Below the critical value of $g$ we show a pair of eigenvalues that 
are non-degenerate and real [green and blue curves
in Fig 3(a)]. At the critical value of $g$ the two real eigenvalues degenerate. Above the critical 
value of $g$ the two eigenvalues become a complex conjugate pair. Thus their real parts 
remain degenerate [green curve in Fig 3(a)] but their imaginary parts are equal in magnitude and
opposite in sign [black and red curves in Fig 3(b)]. Thus the real and imaginary parts of this
pair of eigenvalues plotted as a function of $g$ form a pair of complementary pitchforks as shown
in Fig 3. Thus far the morphology of the transition is the same as in the model of Hatano and Nelson
and in PT quantum mechanics. However our transition has another feature by virtue of a symmetry
of our model that if $E$ is an eigenvalue of our model then so is $-E$ [symmetry (b) discussed in the
paragraph below]. Thus it follows that at
every transition two pairs of real eigenvalues will degenerate [blue and green curves and red and
black curves in Fig 3(a)]. Thereafter these two pairs form two conjugate pairs that remain related
by a sign change. As $g$ is increased further additional quartets of real eigenvalues degenerate
and form complementary conjugate pairs (that is, conjugate pairs that differ in sign). 
An impression of these transitions is conveyed by Fig 4 which depicts the imaginary parts
of the eigenvalues as a function of $g$ for a given value of $\phi$.  For the particular flux
shown in Fig 4 the first transition to complex eigenvalues occurs at a non-zero value of
$g$. Thus there is a small range of $g$ values over which all the eigenvalues remain real
even though their reality is no longer guaranteed by Hermiticity. For other values of the
flux however the first transition to complex eigenvalues happens immediately at $g = 0$.

\begin{figure}[h]
\;\;\; \includegraphics[width=0.45\textwidth]{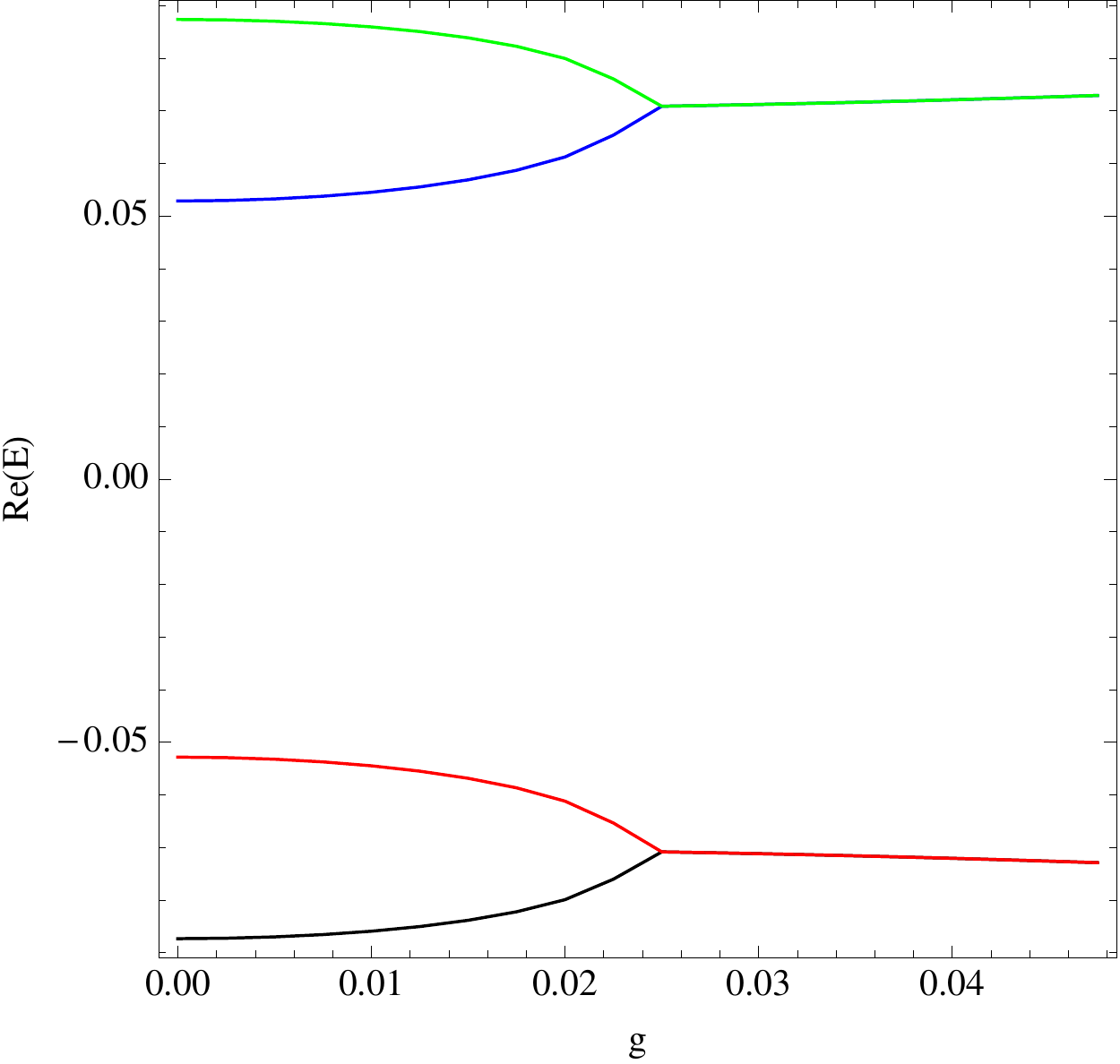}\;\;\;\;
\includegraphics[width=0.45\textwidth]{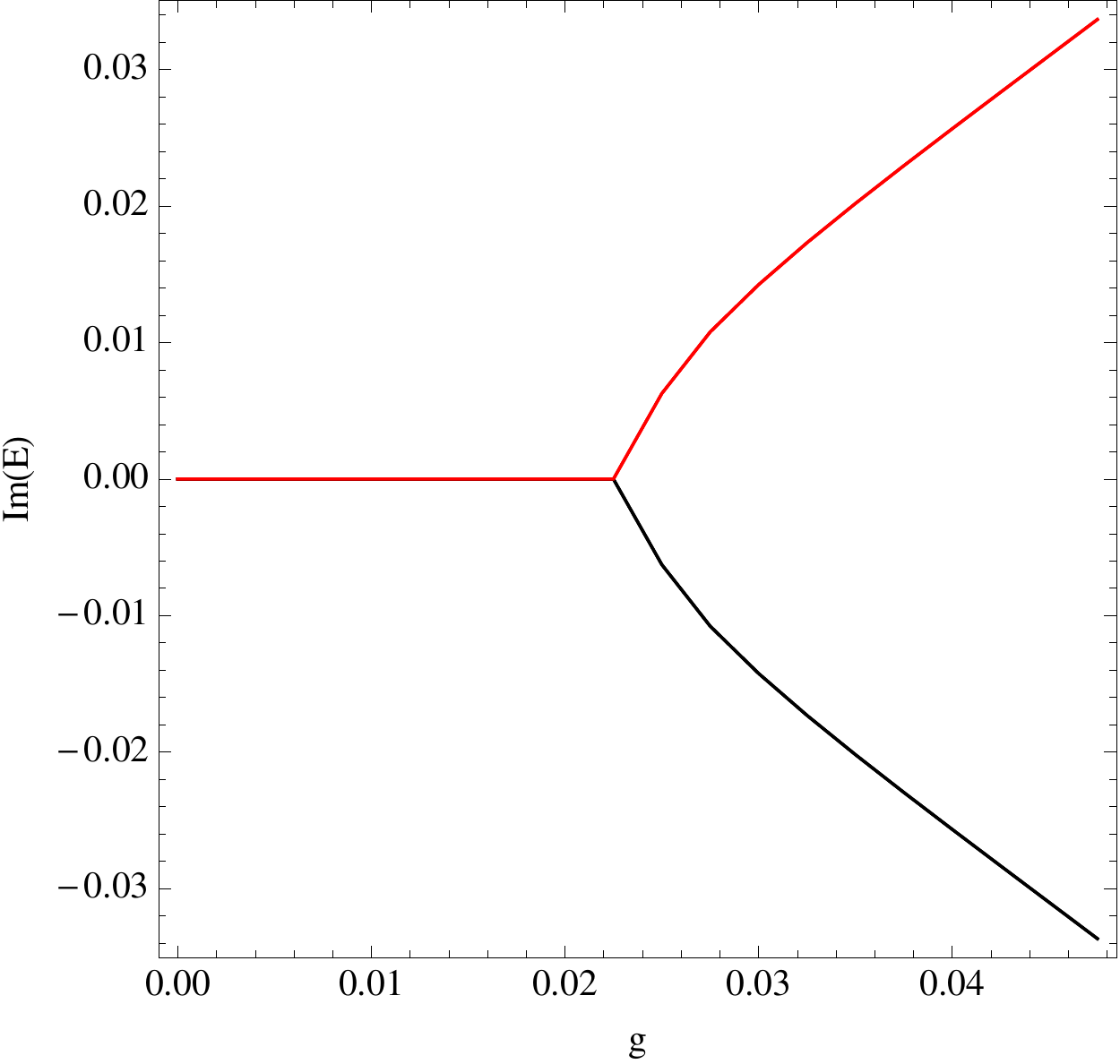}
\caption{The double pitchfork bifurcation. Two pairs of real eigenvalues
degenerate at a critical value of $g$. Thereafter the eigenvalues form
two conjugate pairs that are complementary in the sense that they
differ in sign.
The evolution of the real parts of the four eigenvalues is shown in
(a) and the evolution of the imaginary parts in (b).
In this plot $L = 50$ and $\phi = 0.02$.}
\label{fig:pitchfork}
\end{figure}

\begin{figure}[h]
\begin{center}
\includegraphics[width=0.7\textwidth]{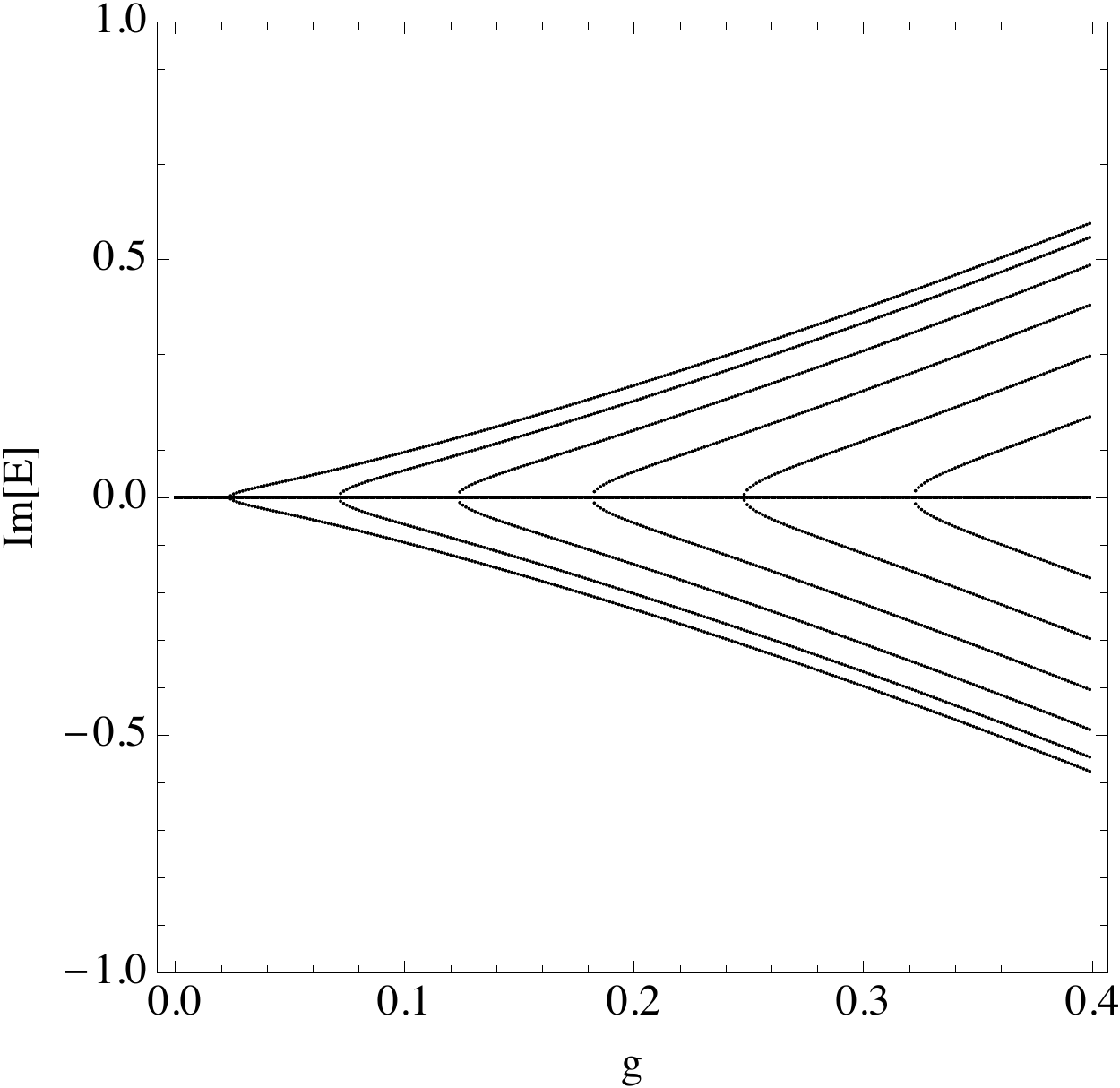}
\end{center}
\caption{Sequence of transitions to complex eigenvalues.
The evolution of the imaginary parts of all the eigenvalues
is shown as a function of $g$ for $L = 50$ and $\phi = 0.02$.}
\label{fig:transitions}
\end{figure}

The Hofstadter butterfly has symmetries that Hofstadter identified and proved\cite{hofstadter}. The spectra we 
calculate have a number of analogous symmetries:  (a) The spectrum is periodic as a function of
the flux $\phi$ with period 1. (b) If $E$ is an eigenvalue for a given flux so is $-E$. (c) The spectrum 
is the same for flux $\phi$ and $1 - \phi$. (d) The spectrum is the same for $g$ and $-g$.
To prove (b) note that if $\xi_m$ is a solution to eq (\ref{eq:cocoon})
with eigenvalue $E$ and wave-vector $k$ then $(-1)^m \xi_m$ with wave-vector $k + \pi$ is a solution to
eq (\ref{eq:cocoon}) but with eigenvalue $-E$. To prove (c) note that if $\xi_m$ is a solution to eq (\ref{eq:cocoon})
with wave-vector $k$ and flux $\phi$ then it is also a solution with wave-vector $-k$ and flux $1 - \phi$ with an
unchanged eigenvalue $E$.  The proofs of (a) and (d) are comparatively simple and are omitted. 

It is not coincidental that the transition to complex eigenvalues has the same double pitchfork form in our
model and in that of Hatano and Nelson as it has in PT quantum mechanics. Recall that in PT quantum
mechanics the Hamiltonian must be PT symmetric {\em i.e.} it must commute with the anti-linear operator $PT$. 
It follows that if $\psi$ is an eigenfunction
of the Hamiltonian with eigenvalue $E$ then $PT \psi$ is an eigenfunction with eigenvalue $E^\ast$. 
Now if eigenstates of the Hamiltonian can be found that are invariant under $PT$ the eigenvalues will
be real. But if the eigenstates are not invariant under $PT$ then $PT$ symmetry is spontaneously broken
and the eigenvalues will be complex \cite{bender2}. The double pitchfork structure emerges from the requirement that
when $PT$ symmetry is broken the eigenvalues must come in conjugate pairs. In the Hatano and Nelson
model and in our model, the role of $PT$ symmetry is played by the anti-linear operation of conjugation.
Inspection of eq (\ref{eq:nelson}) and eq (\ref{eq:cocoon}) shows that if $\xi_m$ is an eigenfunction with
eigenvalue $E$ then $\xi_m^\ast$ is an eigenfunction with eigenvalue $E^\ast$. Thus when the eigenvalues
become complex they must do so in conjugate pairs explaining the double pitchfork structure observed by
us and by Hatano and Nelson. Furthermore the transition to complex eigenvalues is revealed to be a 
manifestation of a spontaneous breaking of conjugation symmetry. Much effort has been devoted to interpreting
delocalization transitions in terms of the concept of spontaneous symmetry breaking \cite{stone}. 
That the delocalization transition of Hatano and Nelson may be regarded as the spontaneous breaking of
conjugation symmetry does not appear to have been remarked upon before.

We conclude with some open questions that deserve further study. 
Design of an electromagnetic analog of our system similar to the
one used to realize Hofstadter's butterfly \cite{microwave} is very desirable as it 
would make our model amenable to experimental study. The recursive
structure of the Hofstadter butterfly was elucidated by Wannier \cite{wannier} who
tracked the evolution of band gaps with flux. We expect the gap trajectories
of the non-Hermitian butterfly 
to be essentially the same as the Hofstadter butterfly 
but the recursive structure of the cocoon would be qualitatively 
different from the butterfly 
and remains to be elucidated. Thouless et al demonstrated
that the bands of the Hofstadter model are characterized
by topological integers \cite{thouless}. The effect of the non-Hermitian perturbation
on these topological invariants is worthy of further investigation.

\end{document}